\documentclass[a4paper,11pt]{article}
\pdfoutput=1 

\usepackage{jheppub} 

\usepackage{mathrsfs}
\usepackage{bbm}
\usepackage{bm}
\usepackage{array,longtable}
\usepackage{multirow}
\usepackage{slashed}
\usepackage[utf8]{inputenc}

\allowdisplaybreaks

\title{\boldmath Disentangling left-handed and right-handed neutrino effects in $B_c \to \tau \bar{\nu}_\tau$ decay}

\author[a,b,c]{Quan-Yi Hu}

\affiliation[a]{Department of Physics, Guangxi Normal University, Guilin 541004, Guangxi, China}
\affiliation[b]{Guangxi Key Laboratory of Nuclear Physics and Technology, Guangxi Normal University, Guilin 541004, Guangxi, China}
\affiliation[c]{School of Physics and Electrical Engineering, Anyang Normal University, Anyang 455000, Henan, China}

\emailAdd{huquanyi@gxnu.edu.cn}

\abstract{Inspired by the anomalies in $B\to D^{(*)} \tau \bar{\nu}_\tau$ decay, in this work, we carefully study the influence of the most general dimension-six effective operators involving right-handed neutrinos on the four cascade decays $B_c \to \tau(\to \nu_\tau h ) \bar{\nu}_\tau $ ($h=\pi,\rho$) and $B_c \to \tau(\to \nu_\tau  \ell \bar{\nu}_\ell) \bar{\nu}_\tau $ ($\ell =\mu, e$). We calculate for the first time the analytical results of the distributions of the differential decay rates of these cascade decays with respect to the energy of the charged particles in the final state. We select a total of 13 new physics benchmark points, among which 9 are derived from the effect of purely left-handed neutrinos, and the other 4 are from the effect of purely right-handed neutrinos. We find that the benchmark points BP4, BP6, BP11, and BP13 can significantly increase the values of these distributions while the BP3 causes a sharp decrease in these distributions. The end-point behavior of the differential distribution $d\Gamma/dE_\pi$ can be used to disentangle the effects of left-handed and right-handed neutrinos. Finally, in order to eliminate the uncertainties brought by the CKM matrix element $V_{cb}$ and the decay constant $f_{B_c}$, we introduce the normalized distributions $d\Gamma/(\Gamma dE_a)$ ($a=\pi,\rho,\mu,e$), which are only sensitive to the right-handed neutrinos. We find that in each normalized distribution there exists a fixed point that is not related to any dimension-six effective operators considered in this work.}

\begin{document} 
\maketitle
\flushbottom

\section{Introduction}
\label{sec:introduction}
The observables $R(D)$ and $R(D^*)$ of $B\to D^{(*)}$ semileptonic decays are theoretically clean and are used to test the universality of lepton flavor between the tau and light leptons. After averaging the measurements of the BaBar collaboration~\cite{BaBar:2012obs,BaBar:2013mob}, the Belle collaboration~\cite{Belle:2015qfa,Belle:2016dyj,Belle:2017ilt,Belle:2019rba}, the Belle II collaboration~\cite{Belle-II:2024ami}, and the LHCb collaboration~\cite{LHCb:2023zxo,LHCb:2023uiv,LHCb:2024jll}, the Heavy Flavor Averaging Group (HFLAV)~\cite{HFLAV:2022esi} showed about $3.31\sigma$ discrepancy~\cite{HFLAV:2024} between the experimental results and the Standard Model (SM) predictions. This discrepancy can be explained by new physics (NP) effects in the $b\to c \tau \nu_\tau$ transition, which is generally described by the following effective Hamiltonian\footnote{It can be derived from the identity $\sigma^{\mu\nu}\gamma_5 = -\frac{i}{2}\epsilon^{\mu\nu\alpha\beta} \sigma_{\alpha\beta}$ that the operators $({\bar c}\sigma^{\mu\nu} P_L b)({\bar\tau}\sigma_{\mu\nu} P_R \nu_\tau) $ and $({\bar c}\sigma^{\mu\nu} P_R b)({\bar\tau}\sigma_{\mu\nu} P_L \nu_\tau) $ are absent. We use the convention $\epsilon_{0123} = - \epsilon^{0123} = 1$.} 
\begin{align}
	\label{eq:Heff}
	\mathcal{H}_\mathrm{eff} = \sqrt{2}G_F V_{cb}\sum_{B=L,R} \big[
	&(g_V^B \bar{c}\gamma^\mu b + g_A^B \bar{c}\gamma^\mu \gamma_5 b)\bar{\tau} \gamma_\mu P_B \nu_\tau \nonumber\\
	&+ (g_S^B \bar{c} b + g_P^B \bar{c} \gamma_5 b)\bar{\tau} P_B \nu_\tau \nonumber\\[3mm]
	&+ g_T^B (\bar{c}\sigma^{\mu\nu} P_B b) \bar{\tau}\sigma_{\mu\nu} P_B \nu_\tau \big] + \mathrm{H.c.},
\end{align}
where $G_F$ is the Fermi constant, $V_{cb}$ is the CKM matrix element, $\sigma^{\mu\nu} = \frac{i}{2}[\gamma^\mu,\, \gamma^\nu]$, and the chirality projectors $P_{L,R} = (1\mp \gamma_5)/2$. The NP effects are encoded in the ten Wilson coefficients $g_i^B$, which are defined at the typical energy scale $\mu_b = m_b$. In the SM, $g_V^L = - g_A^L = 1$ and $g_V^R = g_A^R = g_S^B = g_P^B = g_T^B = 0$. In general, they are complex.

The effective Hamiltonian \eqref{eq:Heff}, which contains both left- and right-handed neutrinos, has been used in phenomenological studies of the $b\to c \tau \nu_\tau$ transitions~\cite{Dutta:2013qaa,Ligeti:2016npd,Iguro:2018qzf,Asadi:2018wea,Greljo:2018ogz,Robinson:2018gza,Azatov:2018kzb,Heeck:2018ntp,Asadi:2018sym,Babu:2018vrl,Bardhan:2019ljo,Shi:2019gxi,Gomez:2019xfw,Mandal:2020htr,Penalva:2021wye,Datta:2022czw,Li:2023gev}. Most of these studies focus on semileptonic decays, such as $B\to D^{(*)} \tau \bar{\nu}_\tau$, $B_c\to J/\psi \tau \bar{\nu}_\tau$, $B_c\to \eta_c \tau \bar{\nu}_\tau$, and $\Lambda_b \to \Lambda_c \tau \bar{\nu}_\tau$. The motivation for introducing right-handed neutrinos is also straightforward, as it would significantly increase the decay rate of the $b\to c \tau \nu_\tau$ transitions, so that the theoretical predicted values of $R(D^{(*)})$ agree with the experimental measurements. Specifically, when $m_{\nu_\tau} \ll E_{\nu_\tau}$, the squared matrix element of a specific decay channel can be represented as
\begin{align}
|\mathcal{M}|^2 = |\mathcal{M}|^2_{\nu_{\tau L}} + |\mathcal{M}|^2_{\nu_{\tau R}} + \mathcal{O}(m_{\nu_\tau} / E_{\nu_\tau}),
\end{align}
where $E_{\nu_\tau}$ the neutrino energy. Neglecting the neutrino mass $m_{\nu_\tau}$, there is no interference between the two neutrino chiralities.

In this work, we take a close look at the leptonic $B_c \to \tau \bar{\nu}_\tau$ decay. In the SM, this channel is sensitive to the CKM matrix element $V_{cb}$ and the decay constant $f_{B_c}$. Unlike the observables in semileptonic decays, which are more or less affected by uncertainties stemming from form factors, we can construct observables that are completely independent of the decay constant $f_{B_c}$ (and of course, the parameter $V_{cb}$ as well) by normalizing the differential decay rate to the total decay rate. If future experiments observe relevant discrepancies, they will undoubtedly be caused by NP beyond the SM. Therefore, the $B_c \to \tau \bar{\nu}_\tau$ decay can serve as an excellent probe to study the NP effects in $b\to c \tau \nu_\tau$ transitions. 

On the other hand, due to angular momentum and parity conservation, only the axial-vector current $g_A^B \bar{c}\gamma^\mu \gamma_5 b$ and pseudoscalar current $g_P^B \bar{c} \gamma_5 b$ in the effective Hamiltonian \eqref{eq:Heff} can contribute to the $B_c$ annihilation process. Furthermore, the equation of motion between matrix elements $\langle 0| \bar{c} \gamma^\mu \gamma_5 b | B_c \rangle \langle \tau \bar{\nu}_\tau|\bar{\tau} \gamma_\mu P_B \nu_\tau|0\rangle$ and $\langle 0| \bar{c} \gamma_5 b | B_c \rangle \langle \tau \bar{\nu}_\tau|\bar{\tau} P_B \nu_\tau|0\rangle$ ensures that the Wilson coefficients $g_A^B$ and $g_P^B$ always appear in the form of combined $g_B \equiv \frac{m_{B_c}^2}{m_\tau (m_b + m_c)}g_P^B - g_A^B\;(B=L,R)$. Accordingly, there exist the relationships $|\mathcal{M}|^2_{\nu_{\tau L}} \propto |g_L|^2$ and $|\mathcal{M}|^2_{\nu_{\tau R}} \propto |g_R|^2$. The chiral enhancement factor in front of the parameters $g_P^B$ makes $B_c \to \tau \bar{\nu}_\tau$ decay highly sensitive to the pseudoscalar sector~\cite{Li:2016vvp,Alonso:2016oyd,Akeroyd:2017mhr}, as predicted by NP models like two-Higgs-doublet models (2HDM)~\cite{Branco:2011iw} or leptoquark models (LQ)~\cite{Buchmuller:1986zs,Dorsner:2016wpm}. We maintain $m_{\nu_\tau} \approx 0$ throughout the whole text. In this way, the contributions from left-handed and right-handed neutrinos will be completely separated. This allows us to conveniently construct observables that are sensitive to right-handed neutrinos.

Experimentally, due to the very short lifetime of the $\tau$, the final $\tau$ does not travel far enough for a displaced vertex. We consider four subsequent $\tau$-decay modes, two hadronic $\tau \to \pi \nu_\tau $ and $\tau \to \rho \nu_\tau$ decays as well as two leptonic $\tau \to \ell \nu_\tau \bar{\nu}_\ell$ ($\ell =\mu, e$) decays. We combine the visible energy distribution of the final-state charged particles and the decay rate of $B_c \to \tau \bar{\nu}_\tau$ to construct observables that are theoretically independent of $V_{cb}$ and the decay constant $f_{B_c}$, and use them to provide a variety of methods to determine whether there is a significant contribution of right-handed neutrinos in the $b\to c \tau \nu_\tau$ transitions.

Our paper is organized as follows. In section~\ref{sec:analytical}, we calculate the analytical results of the measurable energy distributions of the $B_c \to \tau(\to \nu_\tau h ) \bar{\nu}_\tau $ ($h=\pi,\rho$) and $B_c \to \tau(\to \nu_\tau  \ell \bar{\nu}_\ell) \bar{\nu}_\tau $ ($\ell =\mu, e$) decays. The numerical results and discussions are shown in section~\ref{sec:numerical}. Our conclusions are finally made in section~\ref{sec:conclusions}.

\section{Analytical results}
\label{sec:analytical}

\begin{figure}[t]
	\centering
	\includegraphics[width=0.45\textwidth]{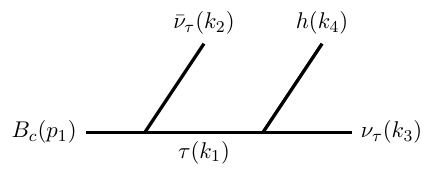}
	\quad
	\includegraphics[width=0.45\textwidth]{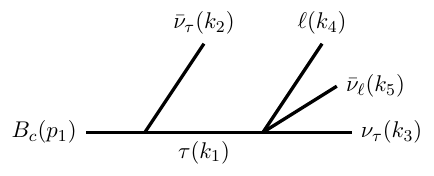}
	\caption{\label{fig:momentum} \small The four-momentum of each particle in $B_c \to \tau(\to \nu_\tau h ) \bar{\nu}_\tau $ ($h=\pi,\rho$) (left) and $B_c \to \tau(\to \nu_\tau  \ell \bar{\nu}_\ell) \bar{\nu}_\tau $ ($\ell =\mu, e$) (right) decays.}
\end{figure}

In this section, within the framework of the effective Hamiltonian \eqref{eq:Heff}, we calculate the analytical expressions for the decay rate of the $B_c \to \tau \bar{\nu}_\tau$ decay, as well as the distributions of differential decay rates with respect to the energy of the final-state charged particles in the $B_c \to \tau(\to \nu_\tau h ) \bar{\nu}_\tau $ ($h=\pi,\rho$) and $B_c \to \tau(\to \nu_\tau  \ell \bar{\nu}_\ell) \bar{\nu}_\tau $ ($\ell =\mu, e$) decays. These four $\tau$-decay modes account for more than 70\% of the total $\tau$-decay width~\cite{ParticleDataGroup:2022pth}. Since no deviation has been observed in experiments for these $\tau$ decays, as in previous literatures (such as~\cite{Ligeti:2016npd,Hu:2020axt,Hu:2021emb,Penalva:2021wye,Li:2023gev}), we assume here that there is no NP effect in $\tau \to \nu_\tau h$ ($h=\pi,\rho$) and $\tau \to \nu_\tau  \ell \bar{\nu}_\ell$ ($\ell =\mu, e$) decays. The definition of the four-momentum of each particle in these two cascade decay modes is shown in figure~\ref{fig:momentum}. The most important results in this section are respectively shown in eqs.~\eqref{eq:dGamm2Epi}, \eqref{eq:dGamm2Erho}, \eqref{eq:dGamm2Elr1}, and \eqref{eq:dGamm2Elr2}.

\subsection{The decay rate of the $B_c \to \tau \bar{\nu}_\tau$ decay}
\label{subsec:tdwBctv}

By inserting the effective Hamiltonian \eqref{eq:Heff} between the initial state $|B_c\rangle$ and the final state $\langle \tau \bar{\nu}_\tau|$, we can obtain the amplitude of this process as
\begin{align}
\label{eq:ampBctv}
\mathcal{M}(B_c \to \tau \bar{\nu}_\tau) = - i\sqrt{2} G_F V_{cb} f_{B_c} m_\tau \bar{u}(k_1) (g_L P_L + g_R P_R ) v(k_2), 
\end{align}
where $g_B \equiv \frac{m_{B_c}^2}{m_\tau (m_b + m_c)}g_P^B - g_A^B$ with $B=L,R$. $m_b$ and $m_c$ are the bottom- and charm-quark running masses in the $\overline{\mathrm{MS}}$ scheme evaluated at the scale $\mu_b$. $f_{B_c}$ is the decay constant of the $B_c$ meson, defined as 
\begin{align}
\label{eq:dcBc}
\langle 0| \bar{c} \gamma^\mu \gamma_5 b | B_c \rangle &\equiv i p_1^\mu f_{B_c}.
\end{align}
Thus the squared matrix element is 
\begin{align}
\label{eq:amp2Bctv}
\left|\mathcal{M}(B_c \to \tau \bar{\nu}_\tau)\right|^2 = 4 G_F^2 |V_{cb}|^2 f_{B_c}^2 m_\tau^2 (k_1\cdot k_2) (|g_L|^2 + |g_R|^2), 
\end{align}
where $k_1\cdot k_2 = \frac{1}{2}\left(m_{B_c}^2 - m_\tau^2 \right)$.

In the $B_c$ rest frame, the two-body phase space is trivial, and for this decay it is given by
\begin{align}
\label{eq:ps2}
\int d\Pi_2 (p_1; k_1, k_2) = \frac{|{\bm k}_2|}{4\pi m_{B_c}} = \frac{m_{B_c}^2 - m_\tau^2}{8\pi m_{B_c}^2}.
\end{align}

Based on eqs.~\eqref{eq:amp2Bctv} and \eqref{eq:ps2}, we can quickly obtain the decay rate of the $B_c \to \tau \bar{\nu}_\tau$ decay as follows
\begin{align}
\label{eq:dwBctv}
\Gamma(B_c \to \tau \bar{\nu}_\tau) = \Gamma(B_c \to \tau \bar{\nu}_\tau)_\mathrm{SM} \left(|g_L|^2 + |g_R|^2\right),
\end{align}
where the result in the SM is 
\begin{align}
\Gamma(B_c \to \tau \bar{\nu}_\tau)_\mathrm{SM} = \frac{G_F^2 |V_{cb}|^2 f_{B_c}^2 m_\tau^2 \left(m_{B_c}^2 - m_\tau^2 \right)^2}{8\pi m_{B_c}^3}.
\end{align}
Our results are consistent with those in refs.~\cite{Gershtein:1976mv,Khlopov:1978id,Hu:2018veh,Mandal:2020htr}.

\subsection{The analytical results in $B_c \to \tau(\to \nu_\tau h ) \bar{\nu}_\tau $ decay mode }
\label{subsec:anareshv}

The squared matrix element of the $B_c \to \tau(\to \nu_\tau h ) \bar{\nu}_\tau $ cascade decay can be written in the narrow width approximation as 
\begin{align}
\label{eq:Mh2}
|\mathcal{M}_h|^2 = \left|\sum_{\lambda_\tau} \frac{\mathcal{M}(B_c \to \tau \bar{\nu}_\tau)\mathcal{M}(\tau \to h \nu_\tau)}{k_1^2 - m_\tau^2 + i m_\tau \Gamma_\tau}\right|^2.
\end{align}
Here $\lambda_\tau$ and $\Gamma_\tau$ are respectively the helicity and total width of intermediate state $\tau$. The helicity $\lambda_\tau$ should be summed before taking the modulus. For $h=\pi$,
\begin{align}
\label{eq:Mpi}
\mathcal{M}(\tau \to \pi \nu_\tau) = i\sqrt{2}G_{F}V_{ud}^{*}f_{\pi}\bar{u}(k_3) \slashed{k}_4 P_{L} u(k_1),
\end{align}
where $V_{ud}^{*}$ is the complex conjugate of CKM matrix element $V_{ud}$, $f_\pi$ is the decay constant of pseudoscalar meson $\pi$ and is defined similarly to $f_{B_c}$ in eq.~\eqref{eq:dcBc}. For $h=\rho$,
\begin{align}
\label{eq:Mrho}
\mathcal{M}(\tau \to \rho \nu_\tau) = \sqrt{2}G_{F}V_{ud}^{*} f_{\rho} m_\rho \bar{u}(k_3) \slashed{\epsilon}^* P_{L} u(k_1), 
\end{align} 
where $\epsilon^\mu$ and $f_{\rho}$ are the polarization vector and decay constant of vector meson $\rho$. Their definitions are as follows
\begin{align}
\label{eq:dcrho}
\langle 0| \bar{u} \gamma^\mu d | \rho \rangle \equiv f_\rho m_\rho \epsilon^\mu.
\end{align}
After calculation, we can obtain the following results
\begin{align}
\label{eq:Mpi2}
|\mathcal{M}_\pi|^2 = N_0 f_\pi^2 (Y_\pi |g_L|^2 + Y_0 |g_R|^2) \delta(k_1^2-m_\tau^2),
\end{align}
where $N_0 \equiv 4\pi G_F^4 |V_{cb}|^2 |V_{ud}|^2 f_{B_c}^2 m_\tau^3 / \Gamma_\tau$, $Y_h \equiv \left(2 k_2\cdot k_4 +m_h^2\right)m_\tau^2 -m_{B_c}^2 m_h^2$, and $Y_0 \equiv m_\tau^2 \left(m_{B_c}^2 - m_\tau^2 - 2 k_2\cdot k_4\right) $, as well as
\begin{align}
\label{eq:Mrho2}
|\mathcal{M}_\rho|^2 = N_0 f_\rho^2 \left[(Y_\rho |g_L|^2 + Y_0 |g_R|^2) + \frac{2m_\rho^2}{m_\tau^2}(Y_0 |g_L|^2 + Y_\rho |g_R|^2)\right] \delta(k_1^2-m_\tau^2).
\end{align}
The first and second terms within the square bracket originate from the contributions of the $\rho$ meson's longitudinal polarization and transverse polarization, respectively. The contribution from transverse polarization is suppressed by $2m_\rho^2 / m_\tau^2$, and the factor of 2 in front of it is due to the fact that there are two degrees of freedom for transverse polarization. The Dirac delta function $\delta(k_1^2-m_\tau^2)$ comes from the fact that we used the narrow width approximation
\begin{align}
\frac{1}{(k_1^2-m_\tau^2)^2+m_\tau^2\Gamma_\tau^2}=\frac{\pi}{m_\tau\Gamma_\tau}\delta(k_1^2-m_\tau^2).
\end{align}

The three-body phase space originally involves a total of nine variables. The conservation of total energy and momentum of the system allows four of these variables to be expressed in terms of the remaining variables. Since the initial particle $B_c$ is spinless, the system satisfies rotational invariance, which can be used to eliminate three more variables. In the end, there are two independent variables left, and it is natural for us to choose them as $k_1^2$ and $E_h$. In the $B_c$ rest frame,  $k_2\cdot k_4 = m_{B_c} E_h - \frac{1}{2}(m_\tau^2 + m_h^2)$, the three-body phase space of this decay mode is given by\footnote{The on-shell relationship for the intermediate state $\tau$ is guaranteed by $\delta(k_1^2-m_\tau^2)$ in eq~\eqref{eq:Mpi2} or \eqref{eq:Mrho2}, therefore $k_1^2$ is a variable here. For the derivation of eq.~\eqref{eq:ps3}, we recommend readers refer to ref.~\cite{Hu:2020axt}.}
\begin{align}
\label{eq:ps3}
d\Pi_3(p_1; k_2,k_3,k_4) = \frac{dE_h dk_1^2}{(4\pi)^3 m_{B_c}}.
\end{align}
The kinematic range of the variable $E_h$ ($h=\pi,\ \rho$) is given by
\begin{align}
\label{eq:rangeEh}
E_h^{\mathrm{min}}\equiv \frac{m_{B_c}^2 m_h^2 + m_\tau^4}{2 m_{B_c} m_\tau^2} \leq E_h \leq E_h^{\mathrm{max}}\equiv \frac{m_{B_c}^2 + m_h^2}{2 m_{B_c}}.
\end{align}

Based on eqs.~\eqref{eq:Mpi2} and \eqref{eq:ps3}, we can directly obtain the differential decay rate of $B_c \to \tau(\to \nu_\tau \pi ) \bar{\nu}_\tau $ decay with respect to the energy of the final $\pi$ as follows
\begin{align}
\label{eq:dGamm2Epi}
\frac{d\Gamma}{dE_\pi} = &\frac{G_F^2 |V_{cb}|^2 f_{B_c}^2 m_\tau^4 \mathcal{B}(\tau \to \pi \nu_\tau)}{4\pi m_{B_c}^2 (m_\tau^2 - m_\pi^2)^2}
\Bigg[\left( m_{B_c}^2 - m_\tau^2 \right)\left( m_\tau^2 - m_\pi^2 \right) \left(|g_L|^2 + |g_R|^2\right)  \nonumber\\
&+ 4 m_{B_c} m_\tau^2 \left(|g_L|^2-|g_R|^2 \right) \left(E_\pi - \frac{E_\pi^{\mathrm{min}} + E_\pi^{\mathrm{max}}}{2} \right) \Bigg]. 
\end{align}
Similarly, by combining eqs.~\eqref{eq:Mrho2} and \eqref{eq:ps3}, we can get the differential decay rate of $B_c \to \tau(\to \nu_\tau \rho ) \bar{\nu}_\tau $ decay with respect to the energy of the final $\rho$ as follows
\begin{align}
\label{eq:dGamm2Erho}
\frac{d\Gamma}{dE_\rho} = &\frac{G_F^2 |V_{cb}|^2 f_{B_c}^2 m_\tau^4 \mathcal{B}(\tau \to \rho \nu_\tau)}{4\pi m_{B_c}^2 (m_\tau^2 - m_\rho^2)^2 ( m_\tau^2 + 2m_\rho^2 )}
\Bigg[\left( m_{B_c}^2 - m_\tau^2 \right) \left( m_\tau^4 + m_\tau^2 m_\rho^2 - 2 m_\rho^4 \right) \left(|g_L|^2 + |g_R|^2\right)  \nonumber\\
&+ 4 m_{B_c} m_\tau^2 \left(m_\tau^2 - 2 m_\rho^2\right) \left(|g_L|^2 - |g_R|^2 \right) \left(E_\rho - \frac{E_\rho^{\mathrm{min}} + E_\rho^{\mathrm{max}}}{2} \right)  \Bigg]. 
\end{align}
In eqs.~\eqref{eq:dGamm2Epi} and \eqref{eq:dGamm2Erho}, we replace the total width of the $\tau$ with the branching ratio of the corresponding secondary decay. The decay rate of $\tau \to \pi \nu_\tau$ is given by
\begin{align}
\Gamma(\tau \to \pi \nu_\tau) = \frac{G_F^2 |V_{ud}|^2 f_\pi^2 (m^2_\tau - m^2_\pi)^2}{16\pi m_\tau} = \Gamma_\tau \mathcal{B}(\tau \to \pi \nu_\tau),
\end{align}
and the decay rate of $\tau \to \rho \nu_\tau$ is given by
\begin{align}
\Gamma(\tau \to \rho \nu_\tau) = \frac{G_F^2 |V_{ud}|^2 f_\rho^2 (m^2_\tau - m^2_\rho)^2 (m^2_\tau + 2m^2_\rho)}{16\pi m^3_\tau} = \Gamma_\tau \mathcal{B}(\tau \to \rho \nu_\tau).
\end{align}

When integrating the eqs.~\eqref{eq:dGamm2Epi} and \eqref{eq:dGamm2Erho} over the interval \eqref{eq:rangeEh} respectively, the term proportional to $|g_L|^2-|g_R|^2$ will disappear, while the term proportional to $|g_L|^2 + |g_R|^2$ will yield the correct narrow-width approximation result, that is, $\Gamma(B_c \to \tau(\to \nu_\tau h ) \bar{\nu}_\tau) = \Gamma(B_c \to \tau \bar{\nu}_\tau) \mathcal{B}(\tau \to h \nu_\tau)$.

\subsection{The analytical results in $B_c \to \tau(\to \nu_\tau \ell \bar{\nu}_\ell ) \bar{\nu}_\tau $ decay mode }
\label{subsec:anareslvv}

The calculation of the squared matrix element $|\mathcal{M}_\ell|^2$ for the $B_c \to \tau(\to \nu_\tau \ell \bar{\nu}_\ell ) \bar{\nu}_\tau $ cascade decay is similar to that of $|\mathcal{M}_h|^2$ in the subsection~\ref{subsec:anareshv}. With the amplitude
\begin{align}
\label{eq:Mlvv}
\mathcal{M}(\tau \to \nu_\tau \ell \bar{\nu}_\ell) = 2\sqrt{2}G_F \bar{u}(k_3) \gamma^\mu P_L u(k_1) \bar{u}(k_4) \gamma_\mu P_L v(k_5),
\end{align}
we can obtain 
\begin{align}
\label{eq:Mlvv2}
|\mathcal{M}_\ell|^2 = &256\pi G_F^4 |V_{cb}|^2 f_{B_c}^2 \frac{m_\tau}{\Gamma_\tau} \delta(k_1^2 - m_\tau^2) (k_3\cdot k_4) \nonumber\\
&\times \left[m_\tau^2 (k_2\cdot k_5) |g_L|^2 + \left(m_{B_c}^2 k_1\cdot k_5 - m_\tau^2 p_1\cdot k_5\right) |g_R|^2\right].
\end{align}

There are a total of $3 \times 4 - 7 = 5 $ independent variables in the four-body phase space of this decay channel. In addition to the variables $k_1^2$ and $E_\ell$, we select the other three variables respectively as the cosine of the angle between the $\ell$ and the $\tau$, namely $\cos\theta$, and any two independent variables in $d^3{\bm k}_3 d^3{\bm k}_5$.
\begin{align}
\label{eq:ps4}
d\Pi_4(p_1; k_2,k_3,k_4,k_5) 
=& \frac{|{\bm k}_1| d k_1^2}{4\pi m_{B_c}} \frac{|{\bm k}_4| dE_\ell d\cos\theta}{2(2\pi)^3} \nonumber\\
&\times\frac{d^3 {\bm k}_3}{(2\pi)^3 2|{\bm k}_3|} \frac{d^3 {\bm k}_5}{(2\pi)^3 2|{\bm k}_5|} (2\pi)^4 \delta^{(4)}(k_1-k_3-k_4-k_5).
\end{align}
The second row in the above equation can be reduced through the following covariant relationship
\begin{align}
\frac{d^3 {\bm k}_3}{(2\pi)^3 2|{\bm k}_3|}\frac{d^3 {\bm k}_5}{(2\pi)^3 2|{\bm k}_5|} (2\pi)^4 \delta^{(4)}(Q-k_3-k_5)k_3^\alpha k_5^\beta
= \frac{1}{96\pi}(Q^2 g^{\alpha \beta} + 2 Q^\alpha Q^\beta) H[Q^2],
\end{align}
which can be easily derived in the rest frame of $Q$. $H[x]$ represents the unit step function, equal to 0 for $x<0$ and 1 for $x \geq 0$. 

The constraint condition $Q^2 = (k_1 - k_4)^2 \geq 0$ divides the integral region of variables $E_\ell$ and $\cos\theta$ into the following two parts. 
\begin{itemize}
\item Region 1
\begin{align}
m_\ell \leq E_\ell \leq E_{\ell 1}\equiv \frac{m_{B_c}^2 m_\ell^2 + m_\tau^4}{2 m_{B_c} m_\tau^2},
\qquad
-1 \leq \cos\theta \leq 1.
\end{align}
\item Region 2
\begin{align}
E_{\ell 1} \leq E_\ell \leq E_{\ell 2}\equiv \frac{m_{B_c}^2 + m_\ell^2}{2m_{B_c}},
\;
\frac{(m_{B_c}^2 + m_\tau^2)E_\ell - m_{B_c} (m_\tau^2 + m_\ell^2)}{(m_{B_c}^2 - m_\tau^2) \sqrt{E_\ell^2 - m_\ell^2}} \leq \cos\theta \leq 1.
\end{align}
\end{itemize}

By combining eqs.~\eqref{eq:Mlvv2} and \eqref{eq:ps4} and then integrating over the variable $\cos\theta$, we can obtain the differential decay rate of $B_c \to \tau(\to \nu_\tau \ell \bar{\nu}_\ell ) \bar{\nu}_\tau $ decay with respect to the energy of final $\ell$. 

For $m_\ell \leq E_\ell \leq E_{\ell 1}$,
\begin{align}
\label{eq:dGamm2Elr1}
\frac{d\Gamma}{dE_\ell} = &\frac{G_F^2 |V_{cb}|^2 f_{B_c}^2 m_{B_c}^2 \mathcal{B}(\tau \to \nu_\tau \ell \bar{\nu}_\ell)}{3\pi y^4 (1-8x^2+8x^6-x^8-24x^4 \ln x)}  s \left(1-y^2\right)^2 \nonumber\\
&\times \Big\{\Big[9 \left(x^2+1\right) \left(y^2+1\right) y^2 z+2
   x^2 \left(y^4-8 y^2+1\right) y^2 \nonumber\\
&-8 \left(y^4+y^2+1\right) z^2\Big]\left(|g_L|^2 + |g_R|^2\right) \nonumber\\
&-  \left(1-y^2\right) \left[x^2 y^2 \left(2 y^2+9 z+2\right)+z \left(y^2 (3-8
   z)-8 z\right)\right] \left(|g_L|^2 - |g_R|^2\right) \Big\}.
\end{align}
To make the expressions more compact, we have defined the following dimensionless parameters
\begin{align}
x \equiv \frac{m_\ell}{m_\tau},\
y \equiv \frac{m_\tau}{m_{B_c}}, \
z \equiv \frac{E_\ell}{m_{B_c}},\
s \equiv \sqrt{z^2 - x^2 y^2}.
\end{align}

For $E_{\ell 1} \leq E_\ell \leq E_{\ell 2}$,
\begin{align}
\label{eq:dGamm2Elr2}
\frac{d\Gamma}{dE_\ell} = &\frac{G_F^2 |V_{cb}|^2 f_{B_c}^2 m_{B_c}^2 \mathcal{B}(\tau \to \nu_\tau \ell \bar{\nu}_\ell)}{12\pi y^4 (1-8x^2+8x^6-x^8-24x^4 \ln x)}\nonumber\\ 
&\times \Bigg\{
\left(y^2-1\right) \Big[x^2 \Big(y^8 (4 s+12 z-9)+9 y^6 (2 (s+z) (z-2)+3) \nonumber\\
&+9 y^4 (4 s-4 z-1)+2 y^2 (3 z (3 z+2)-s (9 z+2))\Big) \nonumber\\
&-y^6 (2 z (8 z-9)
   (s+z)+5)+18 y^2 z (z-s)+16 z^2 (s-z) \nonumber\\
&-5 x^6 y^6-9 x^4 y^4 \left(y^4-3 y^2+1\right)\Big] \left(|g_L|^2 + |g_R|^2\right)  \nonumber\\
&+ \Big[2 s \left(y^2-1\right)^3 \left(x^2 y^2 \left(2 y^2+9 z+2\right)+z \left(3
   y^2-8 \left(y^2+1\right) z\right)\right) \nonumber\\
&+6 y^2 z^2 \left(3 x^2+1\right) \left(y^2+1\right)
   \left(y^4-4 y^2+1\right) \nonumber\\
&+12 x^2 y^2 z \left(\left(4 x^2+6\right) y^4+y^8-2 y^6-2
   y^2+1\right) \nonumber\\
&+y^4 \left( \left(1-5 x^6 \right) y^2 \left(y^2+1\right)- 3 x^2(3 x^2 + 1) \left(y^6+1\right)\right) \nonumber\\
&-16 \left(y^8-2 y^6-2 y^2+1\right) z^3\Big] \left(|g_L|^2 - |g_R|^2\right)
\Bigg\}.
\end{align}

We replace the total width of the $\tau$ with the branching ratio of the $\tau \to \nu_\tau \ell \bar{\nu}_\ell $ decay, whose decay rate is given by
\begin{align}
\Gamma(\tau \to \nu_\tau \ell \bar{\nu}_\ell ) = \frac{G_F^2 m_\tau^5 (1 - 8 x^2 + 8 x^6 - x^8 - 24 x^4 \ln x)}{192\pi^3} = \Gamma_\tau \mathcal{B}(\tau \to \nu_\tau \ell \bar{\nu}_\ell).
\end{align}

Calculating the integral of the variable $E_\ell$ in eqs.~\eqref{eq:dGamm2Elr1} and \eqref{eq:dGamm2Elr2} over the corresponding intervals $m_\ell \leq E_\ell \leq E_{\ell 1}$ and $E_{\ell 1} \leq E_\ell \leq E_{\ell 2}$, respectively, and then summing up the results of these integrals, we also find that the term proportional to $|g_L|^2 - |g_R|^2$ disappears, while the term proportional to $|g_L|^2 + |g_R|^2$ will achieve the correct narrow-width approximation result, namely $\Gamma(B_c \to \tau(\to \nu_\tau \ell \bar{\nu}_\ell ) \bar{\nu}_\tau ) = \Gamma(B_c \to \tau \bar{\nu}_\tau) \mathcal{B}(\tau \to \nu_\tau \ell \bar{\nu}_\ell)$.

\section{Numerical results and discussions}
\label{sec:numerical}

In this section, we present the numerical results corresponding to eqs.~\eqref{eq:dGamm2Epi}, \eqref{eq:dGamm2Erho}, \eqref{eq:dGamm2Elr1} and \eqref{eq:dGamm2Elr2} in various selected NP benchmark points one by one. Additionally, we discuss the observables $d\Gamma/(\Gamma dE_a)$ ($a=\pi,\rho,\mu,e$) that deviate from the predictions of the SM only when right-handed neutrinos exist.

\subsection{The NP benchmark points}
\label{subsec:NPBPs}

The model-independent analyses of the NP effects in $b \to c \tau \nu_\tau$ transitions have been accomplished in numerous literatures~\cite{Alok:2017qsi,Hu:2018veh,Murgui:2019czp,Blanke:2018yud,Shi:2019gxi,Mandal:2020htr,Iguro:2024hyk}. In order to demonstrate the role of disentangling the effect of left-handed and right-handed neutrinos in $B_c \to \tau \bar{\nu}_\tau$ decay by using the differential distributions in eqs.~\eqref{eq:dGamm2Epi}, \eqref{eq:dGamm2Erho}, \eqref{eq:dGamm2Elr1} and \eqref{eq:dGamm2Elr2}, we select various best-fit values as the NP benchmark points. These values are typically performed on a set of chiral base (such as eq.~(2.1) in ref.~\cite{Mandal:2020htr}), which is equivalent to eq.~\eqref{eq:Heff} through the following relationships
\begin{align}
g_V^B &= (\delta_{LB}+C^V_{LB}) + C^V_{RB}, &
g_A^B &= C^V_{RB} - (\delta_{LB}+C^V_{LB}),&\nonumber\\
g_S^B &= C^S_{LB} + C^S_{RB}, &
g_P^B &= C^S_{RB} - C^S_{LB},&
g_T^B = 2C^T_{BB}.
\end{align}
According to the following steps, we respectively select nine NP benchmark points from ref.~\cite{Iguro:2024hyk}, all of which originate from the purely left-handed neutrino contributions; and four NP benchmark points from ref.~\cite{Mandal:2020htr}, all of which originate from the purely right-handed neutrino contributions.

Assuming that the Wilson coefficients $C^{V,S}_{BL}$ are all real numbers and only one is turned on at a time, there are four NP benchmark points as follows~\cite{Iguro:2024hyk} 
\begin{align}
\mathrm{BP1:} \qquad C^V_{LL} &= 0.079, \\
\mathrm{BP2:} \qquad C^V_{RL} &= -0.070, \\
\mathrm{BP3:} \qquad C^S_{LL} &= 0.165, \\
\mathrm{BP4:} \qquad C^S_{RL} &= 0.182. 
\end{align}
Once they are allowed to be complex, we consider the following two additional NP benchmark points that can improve the fitting results~\cite{Iguro:2024hyk} 
\begin{align}
\mathrm{BP5:} \qquad C^V_{RL} &= 0.01 \pm 0.41i, \\
\mathrm{BP6:} \qquad C^S_{LL} &= -0.57 \pm 0.86i.
\end{align}
The last three purely left-handed neutrino benchmark points are derived from the specific LQ hypotheses (which have been evolved to the energy scale $\mu_b$)~\cite{Iguro:2024hyk}
\begin{align}
\mathrm{BP7:} \qquad C^V_{LL} &= 0.07,\ C^S_{RL} = 0.02,\ \mathrm{ for\ the\ singlet\ vector\ LQ\ } U_1,\\
\mathrm{BP8:} \qquad C^V_{LL} &= 0.07,\ C^S_{LL} = \pm 0.15i,\ \mathrm{ for\ the\ singlet\ scalar\ LQ\ } S_1,\\
\mathrm{BP9:} \qquad C^V_{RL} &= \pm 0.50i,\ C^S_{LL} = 0.03 \mp 0.18i,\ \mathrm{ for\ the\ doublet\ scalar\ LQ\ } R_2.
\end{align}

\begin{table}[t]
\tabcolsep 0.15in
\renewcommand\arraystretch{1.2}
\begin{center}
\caption{\label{tab:gLgR} \small The values of $|g_L|$, $|g_R|$ and $|g_L|^2 + |g_R|^2$ at various NP benchmark points. }
\vspace{0.18cm}
\begin{tabular}{cccccccc} 
\hline
      & BP1 & BP2 & BP3 & BP4 & BP5 & BP6 & BP7 
\\ \hline
$|g_L|$ & 1.079 & 1.070 & 0.284 & 1.790 & 1.072 & 5.097 & 1.157
\\ 
$|g_R|$ & 0 & 0 & 0 & 0 & 0 & 0 & 0
\\ 
$|g_L|^2 + |g_R|^2$ & 1.164 & 1.145 & 0.081 & 3.203 & 1.148 & 25.98 & 1.338
\\
\hline \hline
      & BP8 & BP9 & BP10 & BP11 & BP12 & BP13 &
\\ \hline
$|g_L|$  & 1.252 & 0.914 & 1 & 1 & 1 & 1 &
\\ 
$|g_R|$ & 0 & 0 & 0.37 & 2.256 & 0.044 & 1.813 &
\\
$|g_L|^2 + |g_R|^2$ & 1.568 & 0.836 & 1.137 & 6.090 & 1.002 & 4.289 &
\\
\hline
\end{tabular}
\end{center}
\end{table}

Finally, we select four purely right-handed neutrino benchmark points from ref.~\cite{Mandal:2020htr} as follows (They have also been evolved to the energy scale $\mu_b$.)
\begin{align}
\mathrm{BP10:} \qquad C^V_{RR} &= 0.37,\ \mathrm{ for\ the\ mediator\ } V^\mu \sim (1,1,-1),\\
\mathrm{BP11:} \qquad C^S_{LR} &= -0.06,\ C^S_{RR} = 0.46,\ \mathrm{ for\ the\ mediator\  } \Phi \sim (1,2,1/2),\\
\mathrm{BP12:} \qquad C^V_{RR} &= 0.39,\ C^S_{LR} = -0.1,\ \mathrm{ for\ the\ singlet\ vector\ LQ\ } U_1, \\
\mathrm{BP13:} \qquad C^S_{LR} &= 0.418,\ \mathrm{ for\ the\ doublet\ vector\ LQ\ } \tilde{V}_2.
\end{align}

Here, the same treatment as in many literatures (such as \cite{Harrison:2020nrv,Boer:2019zmp,Asadi:2020fdo,Alguero:2020ukk,Hu:2021emb,Li:2023gev}) is adopted, that is, only the central value of the best-fit result is taken as the benchmark point to qualitatively discuss the influence of the NP effect. 

In table~\ref{tab:gLgR}, we list the values of $|g_L|$, $|g_R|$ and $|g_L|^2 + |g_R|^2$ respectively at all NP benchmark points. By combining the predicted value $\mathcal{B}(B_c \to \tau \bar{\nu}_\tau)_{\mathrm{SM}}\simeq 2.2\%$, one can easily obtain $\mathcal{B}(B_c \to \tau \bar{\nu}_\tau)$ at various NP benchmark points. Currently, there is no measured value of $\mathcal{B}(B_c \to \tau \bar{\nu}_\tau)$ in experiment. The future Tera-$Z$ machines, such as CEPC~\cite{Zheng:2020ult} and FCC-ee~\cite{Amhis:2021cfy,Zuo:2023dzn}, can directly measure $\mathcal{B}(B_c \to \tau \bar{\nu}_\tau)$ at $\mathcal{O}(1\%)$ level.\footnote{In ref.~\cite{Zheng:2020ult}, leptonic decays $\tau \to \nu_\tau  \mu \bar{\nu}_\mu$ and $\tau \to \nu_\tau  e \bar{\nu}_e$ are considered, while in refs.~\cite{Amhis:2021cfy,Zuo:2023dzn} the authors focus on the hadronic $\tau \to 3\pi \nu_\tau$ decay. } In phenomenological studies, there are three commonly used choices for the upper bound of $\mathcal{B}(B_c \to \tau \bar{\nu}_\tau)$, namely 10\%~\cite{Akeroyd:2017mhr}, 30\%~\cite{Alonso:2016oyd}, and 60\%~\cite{Blanke:2018yud,Aebischer:2021ilm}. The predicted value of $\mathcal{B}(B_c \to \tau \bar{\nu}_\tau)$ is less than 10\% at all NP benchmark points except BP6 and BP11. The BP6 is obtained when selecting the conservative bound 60\%~\cite{Iguro:2024hyk}.

\begin{figure}[t]
	\centering
	\includegraphics[width=0.35\textwidth]{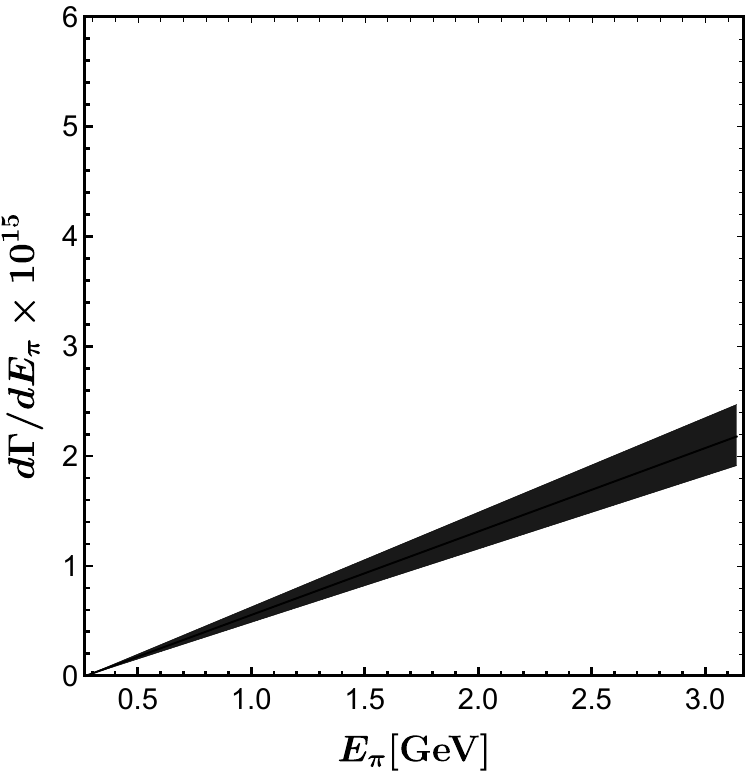}
	\quad
	\includegraphics[width=0.35\textwidth]{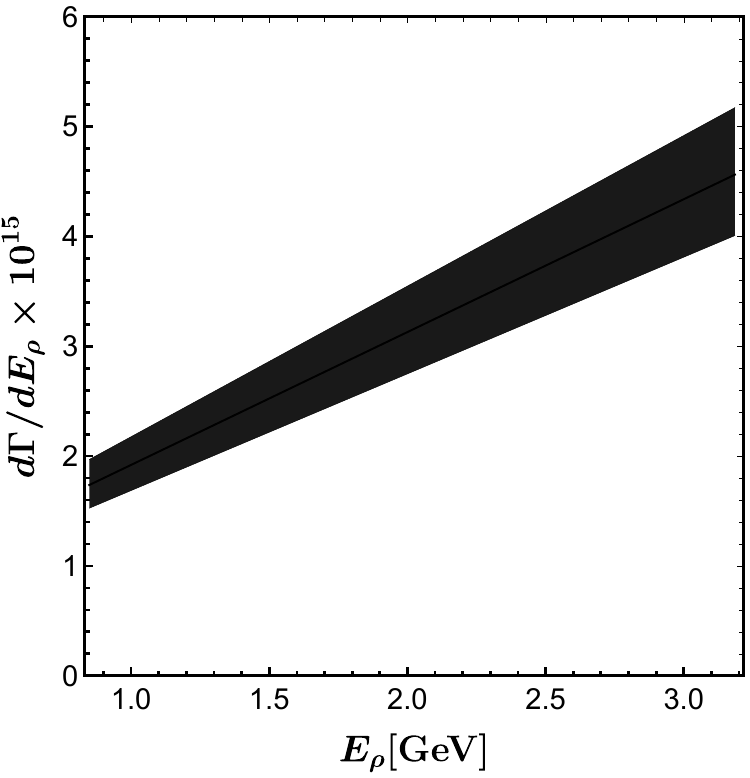}\\
	\includegraphics[width=0.35\textwidth]{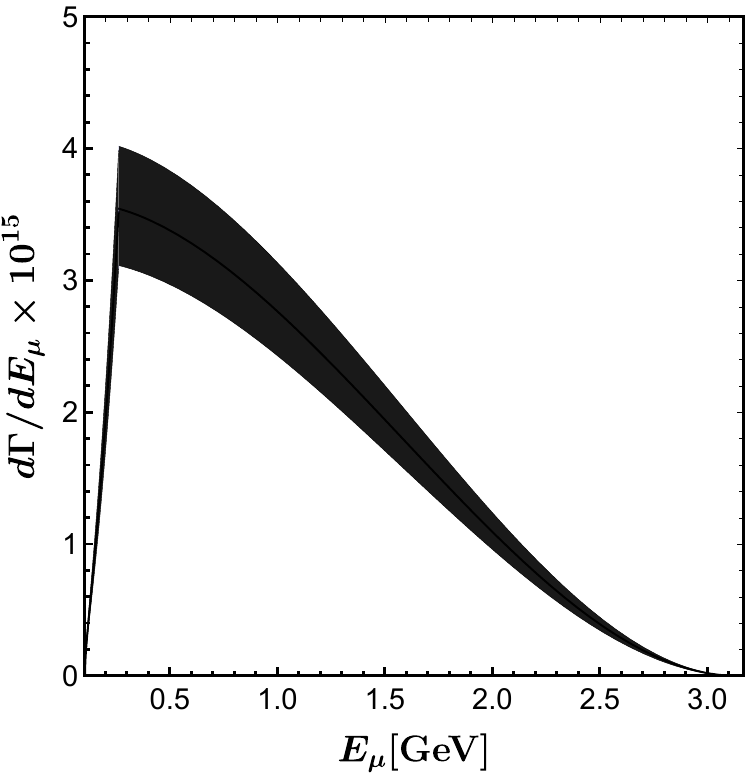}
	\quad
	\includegraphics[width=0.35\textwidth]{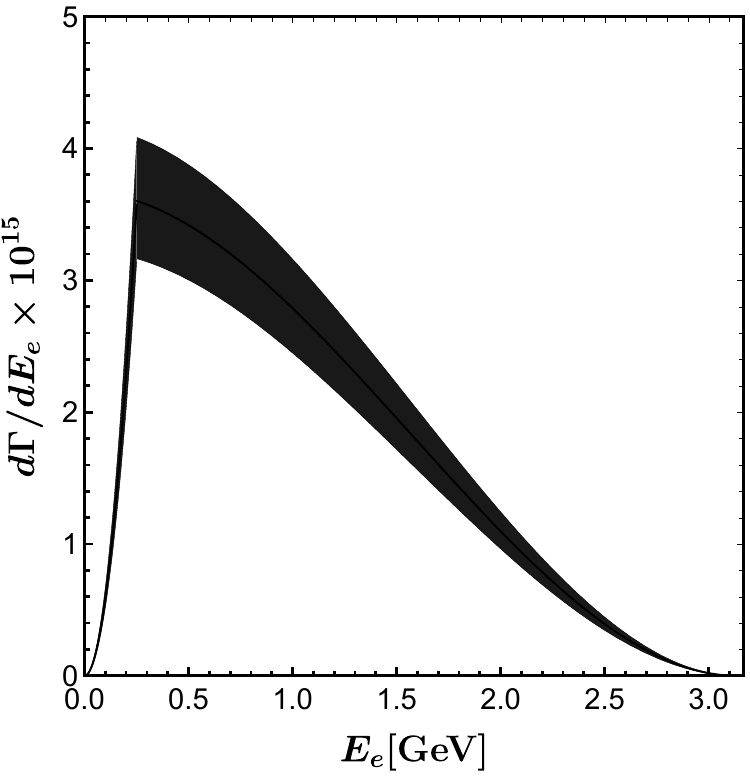}
	\caption{\label{fig:dGam2dEasm} \small The numerical results of differential distributions $d\Gamma/dE_a$ ($a=\pi,\rho,\mu,e$) in the SM. The upper left figure is the distribution of $d\Gamma/dE_\pi$, the upper right figure is the distribution of $d\Gamma/dE_\rho$, the lower left figure is the distribution of $d\Gamma/dE_\mu$, and the lower right figure is the distribution of $d\Gamma/dE_e$.}
\end{figure}
 
\subsection{The numerical results of differential distributions $d\Gamma/dE_a$}
\label{subsec:dGam2dEa}

First of all, within the framework of the SM, we calculate the numerical results of the differential distributions $d\Gamma/dE_\pi$, $d\Gamma/dE_\rho$, $d\Gamma/dE_\mu$, and $d\Gamma/dE_e$ respectively, which are collectively shown in figure~\ref{fig:dGam2dEasm}. The input parameters used in the calculation are all taken from Particle Data Group~\cite{ParticleDataGroup:2022pth}, with the exception of the decay constant $f_{B_c} = 0.434 \pm 0.015$ GeV~\cite{Colquhoun:2015oha}. The uncertainties mainly come from the CKM matrix element $V_{cb}$ and the decay constant $f_{B_c}$.  

The uncertainty of $\pi$ channel is smaller than that of $\rho$ channel because the decay rate of $\pi$ channel is zero at the minimum energy, so the uncertainty is constrained by kinematics; while the $\rho$ channel is nonzero due to the contribution of transverse polarization of $\rho$ meson, thus having a larger uncertainty. By comparing the two subfigures in the bottom row of figure~\ref{fig:dGam2dEasm}, it is not difficult to observe that when the energy of the charged lepton is much greater than its mass, its mass can be safely ignored.

\begin{figure}[t]
	\centering
	\includegraphics[width=0.35\textwidth]{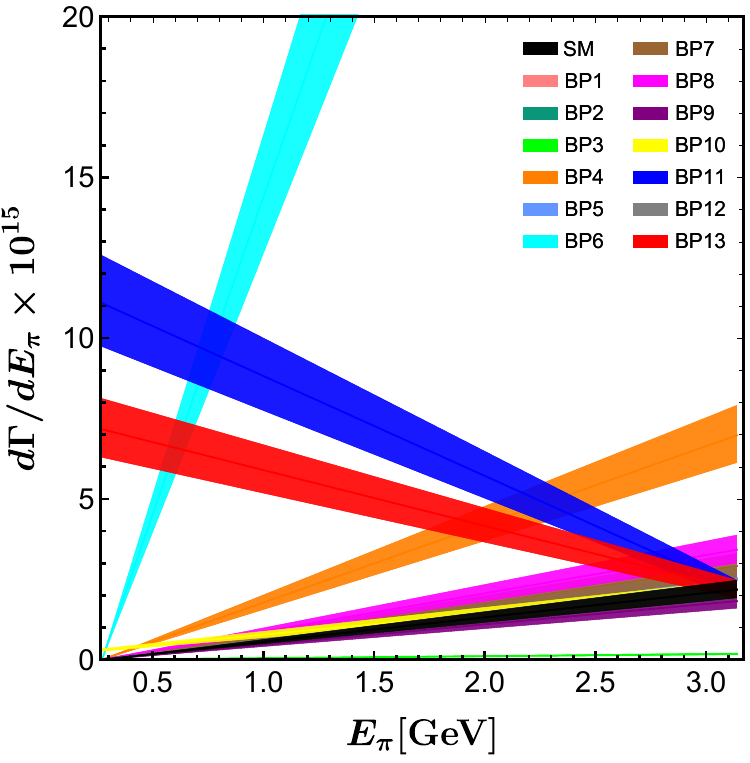}
	\quad
	\includegraphics[width=0.35\textwidth]{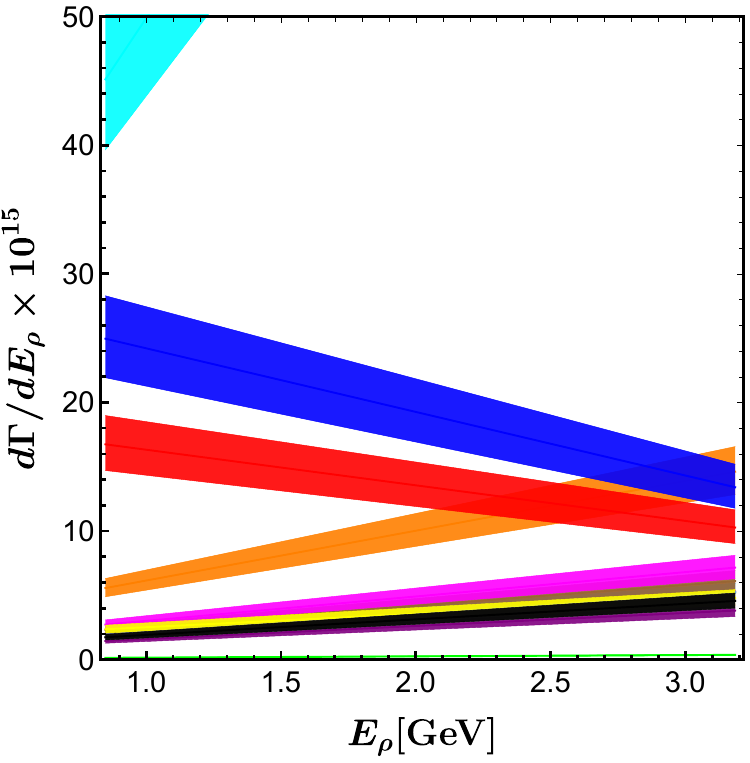}\\
	\includegraphics[width=0.35\textwidth]{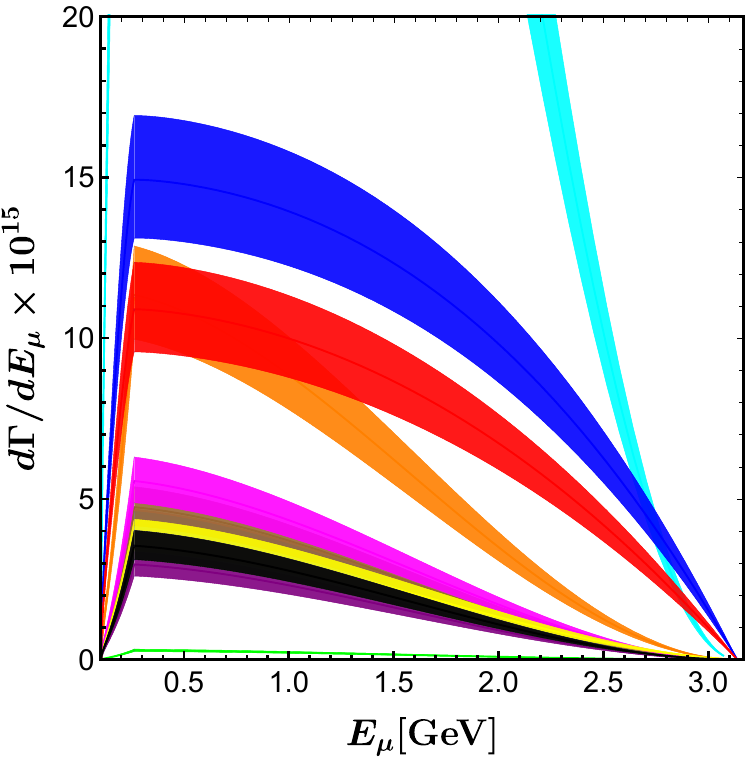}
	\quad
	\includegraphics[width=0.35\textwidth]{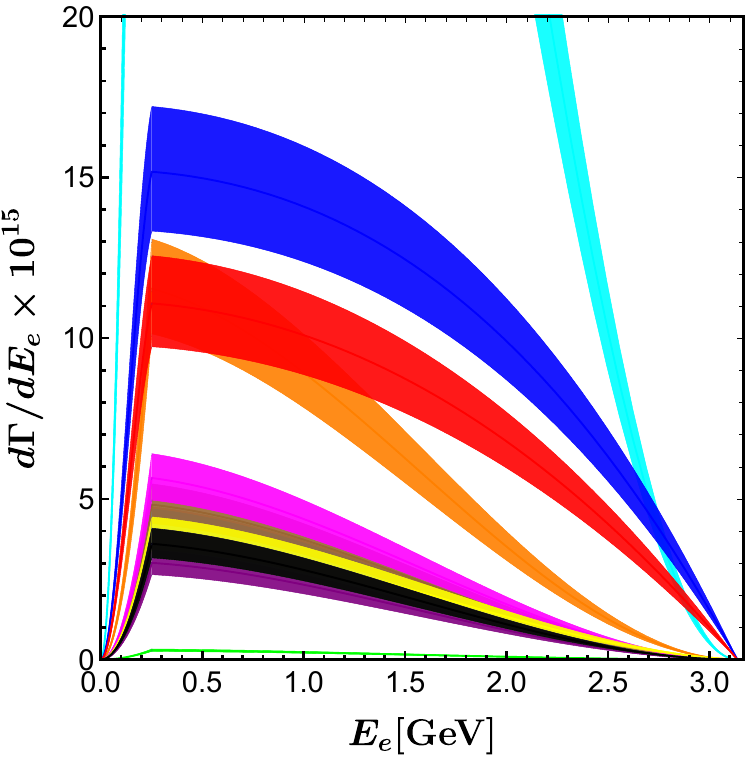}
	\caption{\label{fig:dGam2dEa} \small The numerical results of differential distributions $d\Gamma/dE_a$ ($a=\pi,\rho,\mu,e$) in the NP.}
\end{figure}

NP beyond the SM has the potential to significantly alter the differential distributions shown in figure~\ref{fig:dGam2dEasm}. In figure~\ref{fig:dGam2dEa}, we present the results at various NP benchmark points. Among all the NP benchmark points listed in subsection~\ref{subsec:NPBPs}, only BP3 can lead to a significant decrease in these four differential distributions. The one that causes the most drastic increase in these four differential distributions is BP6. The NP benchmark points BP3 and BP6 actually correspond to exactly the same NP hypothesis, that is to say, there is only one non-zero coefficient $C^S_{LL}$. Such a situation clearly indicates that the differential distribution is extremely sensitive to $C^S_{LL}$. 

In addition, BP4, BP11, and BP13 can also significantly enhance these four differential distributions. Especially BP11 and BP13 have the ability to make the slope of hadronic $d\Gamma/dE_{\pi,\rho}$ negative. Finally, BP7 and BP8 can to some extent increase the four distributions, while the remaining NP benchmark points bring very little influence, which is overshadowed by the uncertainties resulting from the parameters $V_{cb}$ and $f_{B_c}$.

The end-point behavior of the differential distribution $d\Gamma/dE_\pi$ can be used to disentangle the effects of left-handed and right-handed neutrinos. At the minimum energy of meson $\pi$,
\begin{align}
E_\pi &= \frac{m_{B_c}^2 m_\pi^2 + m_\tau^4}{2 m_{B_c} m_\tau^2} = 0.27\ \mathrm{GeV},\\
\frac{d\Gamma}{dE_\pi} &= \frac{G_F^2|V_{cb}|^2 f_{B_c}^2 m_\tau^4 (m_{B_c}^2 - m_\tau^2) \mathcal{B}(\tau \to \pi \nu_\tau)}{2\pi m_{B_c}^2 (m_\tau^2 - m_\pi^2)} |g_R|^2 \nonumber \\
&\simeq  2.18\times 10^{-15} \left(\frac{|V_{cb}|}{0.0411}\right)^2 \left(\frac{f_{B_c}}{0.434 \mathrm{GeV}}\right)^2 |g_R|^2.
\end{align}
Left-handed neutrinos have no contribution to it. Only when right-handed neutrinos exist, this result is non-zero. On the contrary, at the maximum energy of  meson $\pi$,
\begin{align}
E_\pi &= \frac{m_{B_c}^2 + m_\pi^2}{2 m_{B_c}} = 3.14 \ \mathrm{GeV},\\
\frac{d\Gamma}{dE_\pi} &= \frac{G_F^2|V_{cb}|^2 f_{B_c}^2 m_\tau^4 (m_{B_c}^2 - m_\tau^2) \mathcal{B}(\tau \to \pi \nu_\tau)}{2\pi m_{B_c}^2 (m_\tau^2 - m_\pi^2)} |g_L|^2 \nonumber \\
&\simeq  2.18\times 10^{-15} \left(\frac{|V_{cb}|}{0.0411}\right)^2 \left(\frac{f_{B_c}}{0.434 \mathrm{GeV}}\right)^2 |g_L|^2.
\end{align}
Right-handed neutrinos do not interfere with it, and it can be used to detect whether there are other coupling terms of left-handed neutrinos beyond the SM.

In view of the situation that the vector meson $\rho$ presents transverse polarization, it makes it impossible to disentangle the effects of left-handed and right-handed neutrinos at both ends of the differential distribution $d\Gamma/dE_\rho$. At the minimum energy of meson $\rho$, $d\Gamma/dE_\rho \propto |g_R|^2 + (2m_\rho^2/m_\tau^2) |g_L|^2$, and at the maximum energy of meson $\rho$, $d\Gamma/dE_\rho \propto |g_L|^2 + (2m_\rho^2/m_\tau^2) |g_R|^2$. The differential distributions $d\Gamma/dE_{\mu,e}$ have zero values at both ends.

\begin{figure}[t]
	\centering
	\includegraphics[width=0.35\textwidth]{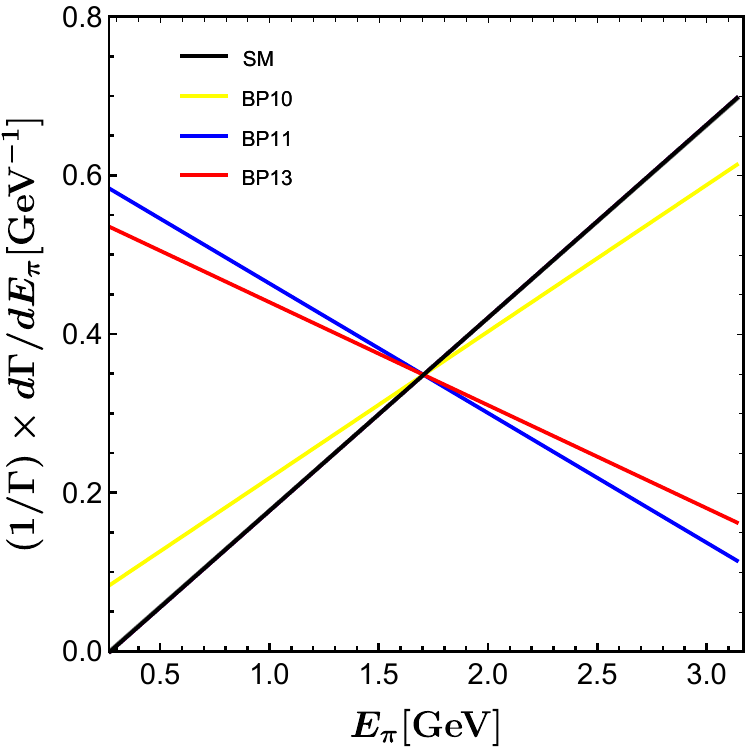}
	\quad
	\includegraphics[width=0.35\textwidth]{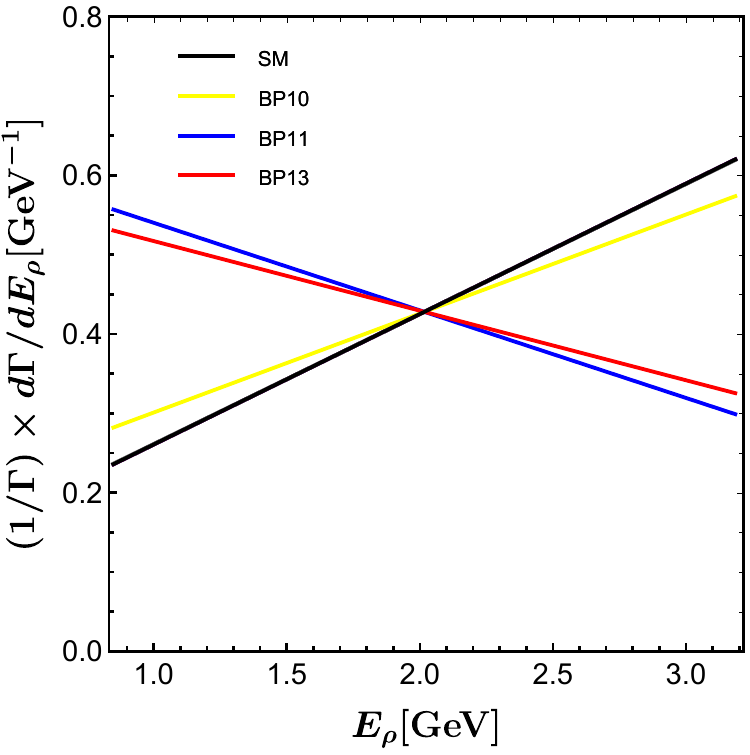}\\
	\includegraphics[width=0.35\textwidth]{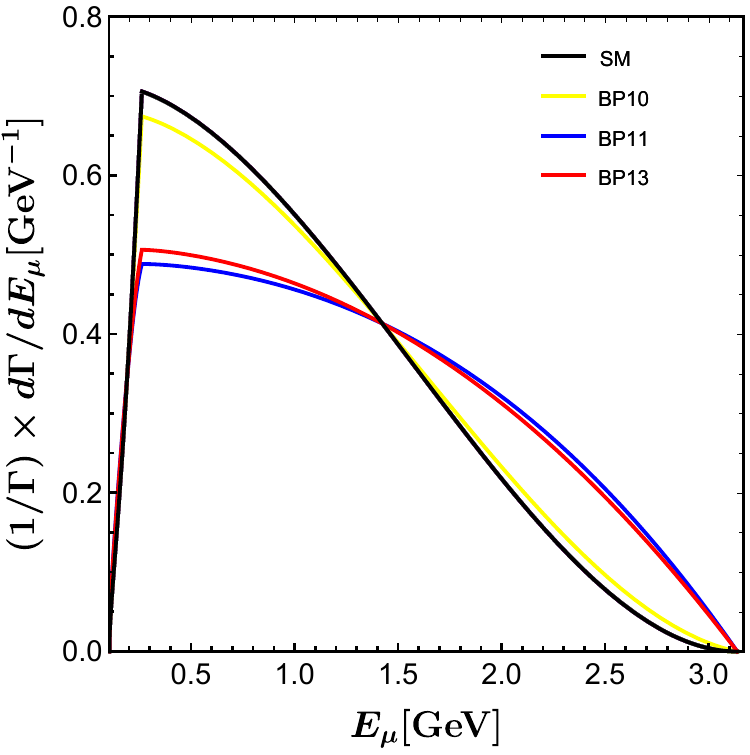}
	\quad
	\includegraphics[width=0.35\textwidth]{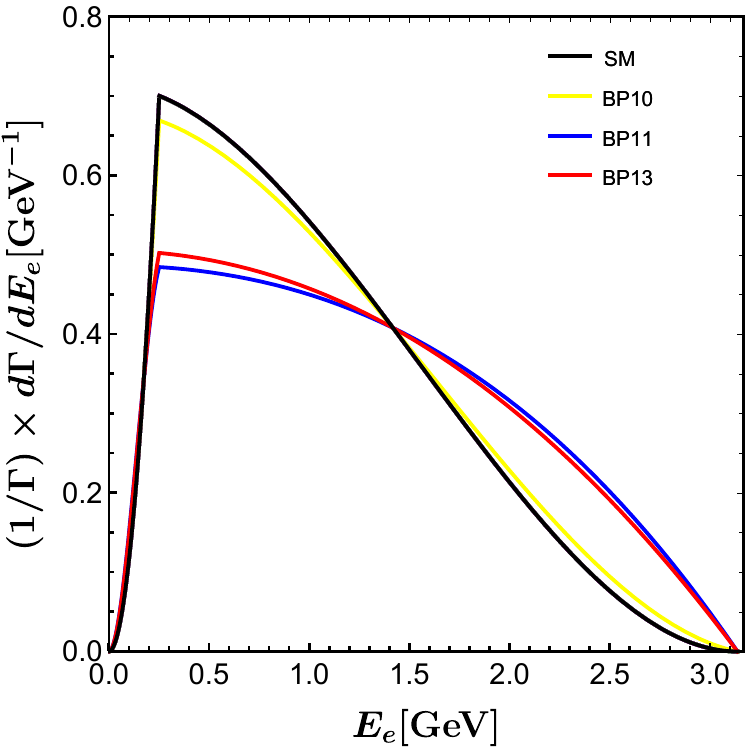}
	\caption{\label{fig:dGam2dEar} \small The numerical results of observables $d\Gamma/(\Gamma dE_a)$ ($a=\pi,\rho,\mu,e$) within the SM as well as in the NP scenarios BP10, BP11 and BP13.}
\end{figure}

\subsection{The observables $d\Gamma/(\Gamma dE_a)$ and the fixed point}
\label{subsec:dGam2GamdEa}

In the subsection~\ref{subsec:dGam2dEa}, it can be clearly observed that the uncertainties caused by the parameters $V_{cb}$ and $f_{B_c}$ interfere with the exploration of NP effects. Therefore, we can construct observables that are completely independent of the parameters $V_{cb}$ and $f_{B_c}$ by dividing the differential distribution of the decay channel by its decay rate. In this subsection, we mainly study the observables $d\Gamma/(\Gamma dE_a)$ ($a=\pi,\rho,\mu,e$), taking $\pi$ channel as an example, which is fully expressed as 
\begin{align}
\frac{d\Gamma}{\Gamma dE_\pi} \equiv \frac{1}{\Gamma(B_c \to \tau(\to \nu_\tau \pi ) \bar{\nu}_\tau)} \frac{d\Gamma(B_c \to \tau(\to \nu_\tau \pi ) \bar{\nu}_\tau)}{d E_\pi}.
\end{align}
Each of these four observables satisfies the relation $\int_{E_{a \mathrm{min}}}^{E_{a \mathrm{max}}}  d\Gamma/(\Gamma dE_a) dE_a = 1$. The analytical expressions presented in eqs.~\eqref{eq:dwBctv}, \eqref{eq:dGamm2Epi}, \eqref{eq:dGamm2Erho}, \eqref{eq:dGamm2Elr1} and \eqref{eq:dGamm2Elr2} clearly indicate that in the absence of right-handed neutrinos, the observables $d\Gamma/(\Gamma dE_a)$ will consistently maintain the predicted values of the SM without any variation. Therefore, we only need to study the effects of purely right-handed neutrino benchmark points BP10 $\sim$ BP13 on the observables $d\Gamma/(\Gamma dE_a)$. We find that the deviation of $d\Gamma/(\Gamma dE_a)$ caused by BP12 is extremely small, and the results of $d\Gamma/(\Gamma dE_a)$ in BP12 is approximately equal to those of the SM.

In figure~\ref{fig:dGam2dEar}, we respectively present the predicted values of observables $d\Gamma/(\Gamma dE_a)$ in the SM and three different NP scenarios BP10, BP11 and BP13. Among the two hadronic decay channels (as detailed in the first row of figure~\ref{fig:dGam2dEar}), the four distribution lines can be well separated, with the two corresponding to BP11 and BP13 standing out particularly. Their slopes are of opposite signs compared to the other two. In the two leptonic decay channels (as detailed in the second row of figure~\ref{fig:dGam2dEar}), the distribution curves corresponding to BP11 and BP13 significantly deviate from those corresponding to the SM and BP10.

More interestingly, we find that there is a fixed point in each distribution $d\Gamma/(\Gamma dE_a)$ ($a=\pi,\rho,\mu,e$) that is independent of any NP effect. For $\pi$ channel, the coordinate of the fixed point is
\begin{align}
E_\pi &= \frac{(m_{B_c}^2 + m_\tau^2)(m_\tau^2 + m_\pi^2)}{4m_{B_c} m_\tau^2} = 1.70\ \mathrm{GeV},\\
\frac{1}{\Gamma}\frac{d\Gamma}{dE_\pi} &= \frac{2 m_{B_c} m_\tau^2}{(m_{B_c}^2 - m_\tau^2)(m_\tau^2 - m_\pi^2)} = 0.35\ \mathrm{GeV}^{-1}.
\end{align}
For $\rho$ channel, the coordinate of the fixed point is
\begin{align}
E_\rho &= \frac{(m_{B_c}^2 + m_\tau^2)(m_\tau^2 + m_\rho^2)}{4m_{B_c} m_\tau^2} = 2.02\ \mathrm{GeV},\\
\frac{1}{\Gamma}\frac{d\Gamma}{dE_\rho} &= \frac{2 m_{B_c} m_\tau^2}{(m_{B_c}^2 - m_\tau^2)(m_\tau^2 - m_\rho^2)} = 0.43\ \mathrm{GeV}^{-1}.
\end{align}
And for $\ell =\mu,e$ channel, the coordinate of the fixed point is
\begin{align}
E_\ell = 1.42\ \mathrm{GeV}, \qquad
\frac{1}{\Gamma}\frac{d\Gamma}{dE_\ell} = 0.41\ \mathrm{GeV}^{-1}.
\end{align}

\section{Conclusions}
\label{sec:conclusions}

Inspired by the $R(D^{(*)})$ anomalies in semileptonic $B \to D^{(*)}$ decays, we carry out a comprehensive study of the leptonic $B_c \to \tau \bar{\nu}_\tau$ decay within the most general dimension-six effective Hamiltonian~\eqref{eq:Heff}, which includes right-handed neutrinos. We further consider four different subsequent decays, namely $\tau \to \pi \nu_\tau$, $\tau \to \rho \nu_\tau$, and $\tau \to \ell \bar{\nu}_\ell \nu_\tau$ ($\ell = \mu ,e$), to construct measurable energy distributions. 

We calculate in sequence the analytical results of the distributions of the differential decay rates of these cascade decays with respect to the energy of the charged particles in the final state. These results are respectively presented in eqs.~\eqref{eq:dGamm2Epi}, \eqref{eq:dGamm2Erho}, \eqref{eq:dGamm2Elr1}, and \eqref{eq:dGamm2Elr2}. Ignoring the mass of the right-handed neutrinos, the contributions from the left-handed and right-handed neutrinos in these results are completely separated and are respectively proportional to the NP parameters $|g_L|^2$ and $|g_R|^2$. Here the parameters $g_{L,R}$ are defined as $g_B \equiv \frac{m_{B_c}^2}{m_\tau (m_b + m_c)}g_P^B - g_A^B\;(B=L,R)$. This enables us to conveniently achieve the decoupling of the effects of left-handed and right-handed neutrinos in the $B_c\to \tau \bar{\nu}_\tau$ decay.

We first show the SM predictions for these four energy distributions. In order to present more effectively the dependence of these distributions on the possible NP effects (especially the NP with only left-handed neutrinos or only right-handed neutrinos), we select a total of 13 NP benchmark points, among which 9 are derived from the effect of purely left-handed neutrinos, and the other 4 are from the effect of purely right-handed neutrinos. Only BP3 can significantly reduce the values of these distributions, while the BP4, BP6, BP11, and BP13 can significantly increase these distributions. We find that at the minimum energy, the $d\Gamma/dE_\pi$ is proportional to $|g_R|^2$ and is nonzero only when the right-handed neutrino exists; while at the maximum energy, $d\Gamma/dE_\pi$ is proportional to $|g_L|^2$. The kinematic phase space of the $B_c \to \tau(\to \nu_\tau \pi ) \bar{\nu}_\tau $ decay is relatively simple and there is no contribution from complex polarization degrees of freedom. Because of this, it stands out among these four cascade  decays.

In order to eliminate the uncertainties introduced by the CKM matrix element $V_{cb}$ and the decay constant $f_{B_c}$, we further consider the distributions $d\Gamma/(\Gamma dE_a)$ ($a=\pi,\rho,\mu,e$). In the absence of right-handed neutrinos, these four distributions always maintain the predicted values of the SM. If deviations from the predicted values of the SM are measured in experiments, it necessarily implies the existence of right-handed neutrinos. Furthermore, we find that in each distribution there exists a fixed point that is not related to any NP in effective Hamiltonian~\eqref{eq:Heff}. If future experimental measurements deviate from these fixed points, it would imply the existence of new physics beyond the Hamiltonian~\eqref{eq:Heff}, potentially hinting at the existence of undiscovered light particles and interactions. We hope that the findings discussed in this work can be tested on future Tera-$Z$ machines such as CEPC and FCC-ee.

\acknowledgments

This work is supported by the National Natural Science Foundation of China under Grant No.~12105002, and the Guangxi Natural Science Foundation under
Grant No.~2023GXNSFB A026270.

\bibliographystyle{JHEP}
\bibliography{ref}

\end{document}